\documentclass[10pt,A4paper,conference]{IEEEtran}

\newtheorem{teo}{Theorem}

\usepackage{graphicx}
\usepackage{amsmath}
\usepackage{amsfonts}
\usepackage{bm}
\usepackage{subfigure}
\begin{document}

\title{Non prefix-free codes for constrained sequences}

\author{\authorblockN{Marco Dalai \& Riccardo Leonardi}
\authorblockA{Department of Electronics for Automation\\
University of Brescia, Italy\\
Email: \{marco.dalai, riccardo.leonardi\}@ing.unibs.it}
}
%

\maketitle

\begin{abstract}
In this paper we consider the use of variable length non prefix-free codes for coding constrained sequences of symbols.
We suppose to have a Markov source where some state transitions are impossible, i.e. the stochastic matrix associated with the Markov chain has some null entries. We show that classic Kraft inequality is not a necessary condition, in general, for unique decodability under the above hypothesis and we propose a relaxed necessary inequality condition.
This allows, in some cases, the use of non prefix-free codes that can give very good performance, both in terms of compression and computational efficiency. Some considerations are made on the relation between the proposed approach and other existing coding paradigms.
\end{abstract}

\section{Introduction}
Variable length codes are usually considered to have to satisfy the Kraft inequality (\cite{kraft-kraftineq}), as it was proved that this is a necessary condition for unique decodability in \cite{mcmillan-kraftineq}. Unique decodability is always defined as a characteristic of the set of code words only, and no reference to the type of source is considered. Thus, imposing unique decodability we are requiring any sequence of symbols being distinguishable. This fact is clearly perfectly acceptable if the source can produce any sequence of symbol with non-null probability as, for example, in the memoryless source case. But what happens if the source is not memoryless and, more precisely, not all sequences are allowed? Consider for example a source with three symbols $A$, $B$ and $C$, and suppose that symbol $A$ can never be followed by $B$. This fact is a characteristic of the source that we can consider known for the encoding problem, as we actually do in the classic Shannon paradigm where no bits are assigned to null-probability events. Thus, the decoder is usually supposed to know that an event is not possible, at least from the fact that that no code word is assigned to that event. Thus, suppose we use the codewords $'0'$, $'1'$ and $'01'$ for our source symbols $A$, $B$ and $C$ respectively. It is easy to see that any sequence of symbols that can be generated by the source is uniquely specified by concatenating the codewords of every single symbol. This is because with this code one may only confuse sequences of symbols with sequences that our source cannot generate.
The objective of this paper is to propose the notion of unique decodability of a set of codewords relatively to a specific information source. We show simple examples of Markov sources that can be efficiently encoded by non-prefix free codes. Our examples also show that there are sources for which a coding technique exists such that the expected number of bits of the code for the first $n$ symbols of the source is strictly smaller than the entropy of those first $n$ symbols (for every $n$). This fact is often erroneously considered impossible, and this error arises from considering the Kraft inequality a necessary condition for any type of sources.

\section{A curious Markov chain example}
\begin{figure}[b]
\centering
\includegraphics[width=4cm]{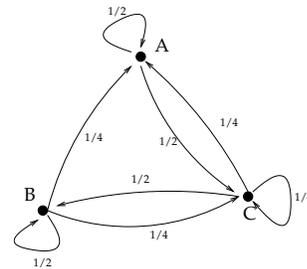}
\caption{Graph, with transition probabilities, for a Markov Chain.}
\label{fig:markov_3}
\end{figure}

Suppose, so as to analyze more throughly the example of the introduction, we have a source generating symbols $X_1,X_2,X_3,\ldots$ extracted from the set $\mathcal{X}=\{A,B,C\}$ following the Markov chain graphically shown in figure \ref{fig:markov_3}.
In formulae, if we call $\mathbf{S}_i$ the probability distribution row vector on $\mathcal{X}$ at step $n$, we have $\mathbf{S}_{i+1}=\mathbf{S}_i\mathbf{P}$, where $\mathbf{P}$ is the transition probability matrix
\begin{equation}
\mathbf{P}=\left[
\begin{array}{ccc}
1/2 & 0 & 1/2\\
1/4 & 1/2 & 1/4\\
1/4 & 1/2 & 1/4
\end{array}
\right]
\end{equation}
If we suppose the initial state has uniform probability $\mathbf{S}_1=[1/3,1/3,1/3]$, it is easy to verify that the process is stationary, i.e $\mathbf{S}_i=\mathbf{S}_1$ for every $i$. Thus, the entropy $H(X_1,X_2,\ldots,X_k)$ of the first $k$ symbols can be easily computed. We have
\begin{multline}
H(X_1,X_2,\ldots,X_k)=\\
H(X_1)+H(X_2|X_1)+\cdots+H(X_k|X_{k-1})=\\
H(X_1)+H(X_2|X_1)(k-1)=\log(3)+\frac{4}{3}(k-1)\nonumber
\end{multline}
Let us consider now the codeword assignment $A\rightarrow 0$, $B\rightarrow 1$, $C\rightarrow 01$. This code is clearly not prefix-free but, as explained in the introduction, when used for this source no ambiguity can arise.
If $l(X)$ is the length of the code associated to $X$, we can compute the expected number of bits in coding the first $k$ symbols as
\begin{multline}
E[l(X_1 X_2 X_3 \cdots X_k)]= E\left[\sum_{i=1}^k l(X_i)\right]=\\
\sum_{i=1}^k E[l(X_i)]=k\left(\frac{1}{3}+\frac{1}{3}+\frac{2}{3}\right)=\frac{4}{3}k\nonumber
\end{multline}
It is easy to verify that this value is strictly smaller than the entropy $H(X_1,X_2,\cdots,X_k)$ above computed.
What happens with our coding procedure? And what happens when the number of symbols goes to infinity?
Looking carefully at our example, we note that our coding strategy uses an expected number of $4/3$ bits for coding the first symbol, while its entropy is $\log 3$. For the following symbols, in turn, the entropies $H(X_2|X_1)$, $H(X_3|X_2)\ldots$ equal $4/3$ bits, and thus they have exactly the same value as the number of bits used by our code.
So, we can say that our code only gains in the first symbol and not substantially (as it is obvious from AEP for ergodic sources). But this fact is somehow interesting; our code assigns to the first symbol a number of bits smaller than its entropy, using the memory properties of the source, without affecting unique decodability. Thus, if we consider the case when only a finite number of symbols are given the problem of finding the optimal coding strategy arises. Furthermore, we should consider that in the more general case, when higher order constraint are eventually present, the problem becomes much more intriguing.

\section{Kraft inequality for constrained sequences}

McMillan showed in \cite{mcmillan-kraftineq} that a necessary condition for the unique decodability of a set of $n$ codewords is that their lengths $l_1,l_2,\ldots,l_n$ satisfy the Kraft inequality
\begin{equation}
\sum_{i=1}^n 2^{-l_i}\leq 1
\end{equation}
Karush (\cite{karush-1961}) gave a simple proof of this fact by considering that for every $k>0$ the following inequality must be satisfied
\begin{equation}\label{karush}
{\left(\sum_i 2^{-l_i}\right)}^k \leq k\, l_{\text{max}}
\end{equation}
The term on the left hand side of (\ref{karush}) can be written as the sum of weights of codes of possible sequences of $k$ symbols. For example, a sequence starting with $x_1,x_3,x_2,\ldots$ gives a term $2^{-l_1}2^{-l_3}2^{-l_2}\cdots$ in the expantion of the left hand side of (\ref{karush}). In order to have only one sequence assigned to every code the above inequality is necessary. But if we only want to distinguish between the possible sequences generated by a constrained source, we may rewrite the condition in a less stringent form. Let us consider once more as an example the source of fig. \ref{fig:markov_3} with $l_1$, $l_2$ and $l_3$ lengths assigned respectively to $A$, $B$ and $C$. Thus, terms on the expansion of left hand side of (\ref{karush}) that contains $\cdots 2^{-l_1}2^{-l_2}\cdots$ should not be considered as $B$ cannot follow $A$ in a source sequence. Let us consider the matrix
\begin{equation}
\mathbf{Q}=\left[
\begin{array}{ccc}
2^{-l_1} & 0 & 2^{-l_1}\\
2^{-l_2} & 2^{-l_2} & 2^{-l_2}\\
2^{-l_3} & 2^{-l_3} & 2^{-l_3}
\end{array}
\right];
\end{equation}
it is possible to verify that the really necessary correspondent  of eq. (\ref{karush}) for our source should be written, for $k>0$, as
\begin{equation}\label{eq:prod_minore}
\left[
\begin{array}{ccc}
1 & 1 &1
\end{array}\right]
\mathbf{Q}^{k-1}
\left[
\begin{array}{c}
2^{-l_1}\\
2^{-l_2}\\
2^{-l_3}
\end{array}
\right]
\leq k\, l_{\text{max}}\
\end{equation}
It is possible to show that a necessary condition for this inequality to be satisfied for every $k$ is that the matrix
$\mathbf{Q}$ has spectral radius\footnote{The spectral radius of a matrix is defined as the greatest modulus of its eigenvalues.} at most equal to 1. We state and prove this fact in the general case.
\begin{teo}
Let $\mathbf{P}$ be an irreducible $n\times n$ stochastic matrix and $\mathbf{l}=[l_1,l_2,\ldots,l_n]$ a vector of $n$ integers. Let $\mathbf{Q}$ be the $n\times n$ matrix such that
\begin{equation}
\mathbf{Q}_{ij}=
\begin{cases}
0 & \mbox{if } {P}_{ij}=0\\
2^{-l_i} & \mbox{if }  {P}_{ij}>0
\end{cases}
\end{equation}
Then, a necessary condition for the codeword lengths $l_1$, $l_2$, \ldots, $l_n$ to be lengths of a uniquely decodable code for a Markov source with transition probability matrix $\mathbf{P}$ is that $\rho(\mathbf{Q})\leq 1$, where $\rho(\mathbf{Q})$ is the spectral radius of $\mathbf{Q}$.
\end{teo}
\begin{proof}
We follow Karush's proof of McMillan theorem. Suppose without loss of generality that the set of our source symbols is $\mathcal{X}=\{1,2,\ldots,n\}$, and call $\mathcal{X}^{(k)}$ the set of all sequences of $k$ symbols that can be produced by the source. Let us set $\mathbf{L}=2^{-\mathbf{l}}=[2^{-l_1},2^{-l_2},\ldots,2^{-l_n}]$ and define, for $k>0$, 
\begin{equation}
\mathbf{V}_k^T=\mathbf{Q}^{k-1}\mathbf{L}^T.
\end{equation}
Then it is easy to see by induction that the $i$-th component of $\mathbf{V}_k$ is written as
\begin{equation}
\mathbf{V}_k^i=\sum_{x_1, x_2, \ldots, x_k}2^{-l_{x_1}-l_{x_2}\cdots -l_{x_k}}
\end{equation}
where the sum runs over all elements $(x_1, x_2,\ldots, x_k)$ of $\mathcal{X}^{(k)}$ with $x_1=i$.
So, if we call $\mathbf{1}_n$ the row vector composed of $n$ 1's, we have
\begin{equation}
\mathbf{1}_n \mathbf{Q}^{k-1}\mathbf{L}^T=\sum_{x_1, x_2, \ldots, x_k}2^{-l_{x_1}-l_{x_2}\cdots -l_{x_k}}
\end{equation}
where the sum now runs over all elements of $\mathcal{X}^{(k)}$. Thus, reindexing the sum with respect to the total length $l=l_{x_1}+l_{x_2}+\cdots+l_{x_k}$ and calling $c_l$ the number of sequences of $\mathcal{X}^{(k)}$ to which correspond a code of length $l$, we have
\begin{equation}
\mathbf{1}_n \mathbf{Q}^{k-1}\mathbf{L}^T=\sum_{l=k l_{\text{min}}}^{k l_{\text{max}}}c_l 2^{-l}
\end{equation}
where $l_{\text{min}}$ and $l_{\text{max}}$ are respectively the maximum and the minimum of the values $l_i,i=1,2,\ldots,n$.
Since the code is uniquely decodable, all $c_l$ sequences of length $l$ must be different and so they are at most $2^l$. This implies that, for every $k>0$, we must have
\begin{equation}\label{eq:disug_powerlinear}
\mathbf{1}_n \mathbf{Q}^{k-1}\mathbf{L}^T\leq \sum_{l=k l_{\text{min}}}^{k l_{\text{max}}}2^{\,l} 2^{-l}=k(l_{\text{max}}-l_{\text{min}}+1)
\end{equation}
Now, note that the irreducible matrix $\mathbf{Q}$ is also nonnegative. Thus, for the Perron-Frobenius theorem (see \cite{minc-book} for details), its spectral radius $\rho(\mathbf{Q})$ is also an eigenvalue\footnote{Note that in general the spectral radius is not an eigenvalue as it is defined as the maximum of $|\lambda|$ over all eigenvalues $\lambda$.}, with algebraic multiplicity 1 and with positive associated eigenvector. As the vectors $\mathbf{1}_n$ and $\mathbf{L}$ are both positive, this implies that the term on the left hand side of eq. (\ref{eq:disug_powerlinear}) asymptotically grows like $\rho(\mathbf{Q})^{k-1}$. On the contrary, the right hand side term only grows linearly with $k$; so, taking the limit as $k\to \infty$ in eq. (\ref{eq:disug_powerlinear}) we conclude that $\rho(\mathbf{Q})\leq 1$.
\end{proof}

We note that if the $\mathbf{P}$ matrix has all strictly positive entries, the matrix $\mathbf{Q}$ reduces to have all equal columns, and its spectral radius is exactly $\sum 2^{-l_i}$. Thus, for non-constrained sequences, we obtain the classic Kraft inequality. Furthermore, as the spectral radius of a nonnegative positive matrix increases if any of the elements increases, we note that the situation $\rho(\mathbf{Q})=1$ is a strong condition on $\mathbf{P}$ and $\mathbf{l}$. In the sense that if for a given matrix $\mathbf{P}$ there is a decodable code with codeword lengths $l_i,i=1,\ldots,n$ such that $\rho(\mathbf{Q})=1$, then there is no decodable code with lengths $l_i'$ if $l_i'<l_i$ for some $i$. Also, it is not possible to remove constraints from the Markov chain while keeping unique decodability property.

The most important remark, however, concerns the non sufficiency of the stated condition. In fact, while the classic Kraft inequality is a necessary and sufficient condition for the existence of a uniquely decodable code for an unconstrained sequence, the found inequality $\rho(\mathbf{Q})\leq 1$ is unfortunately only necessary, and not sufficient. We discuss this point in the next section, where we propose an extension of the Sardinas Patterson test for testing the unique decodability of a code for a constrained sequence.

\section{Non sufficiency and Sardinas Patterson test}
\begin{figure}
\centering
\subfigure[\label{fig:Markov_2}]{\includegraphics[width=3.5cm]{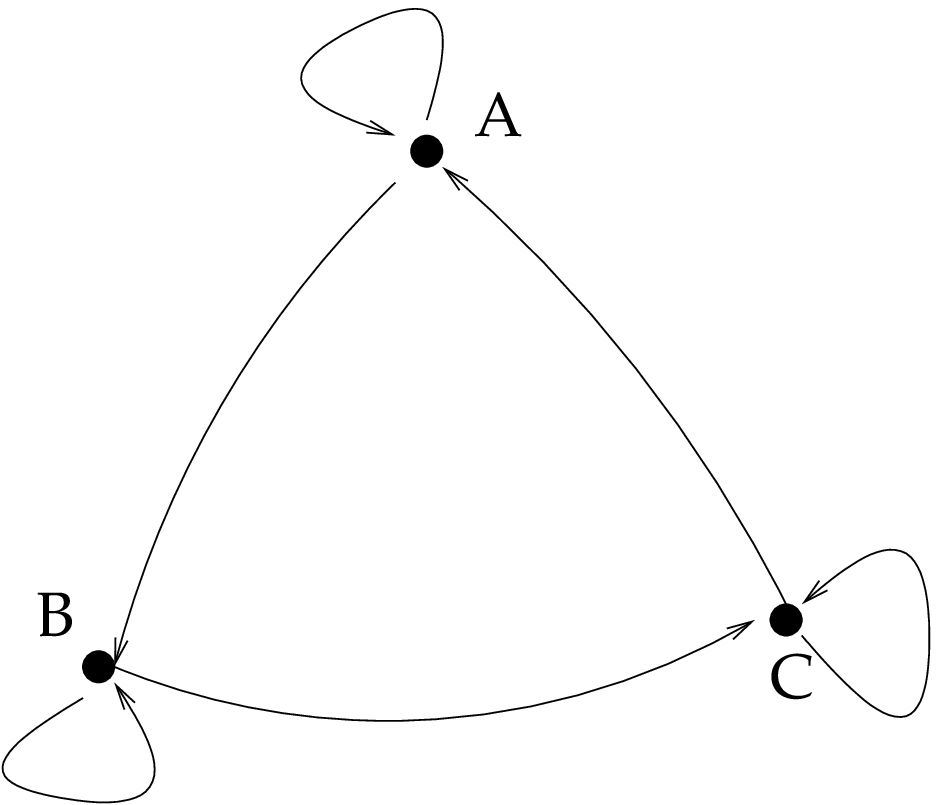}}
\hspace{20pt}
\subfigure[\label{fig:Markov_3}]{\includegraphics[width=3.5cm]{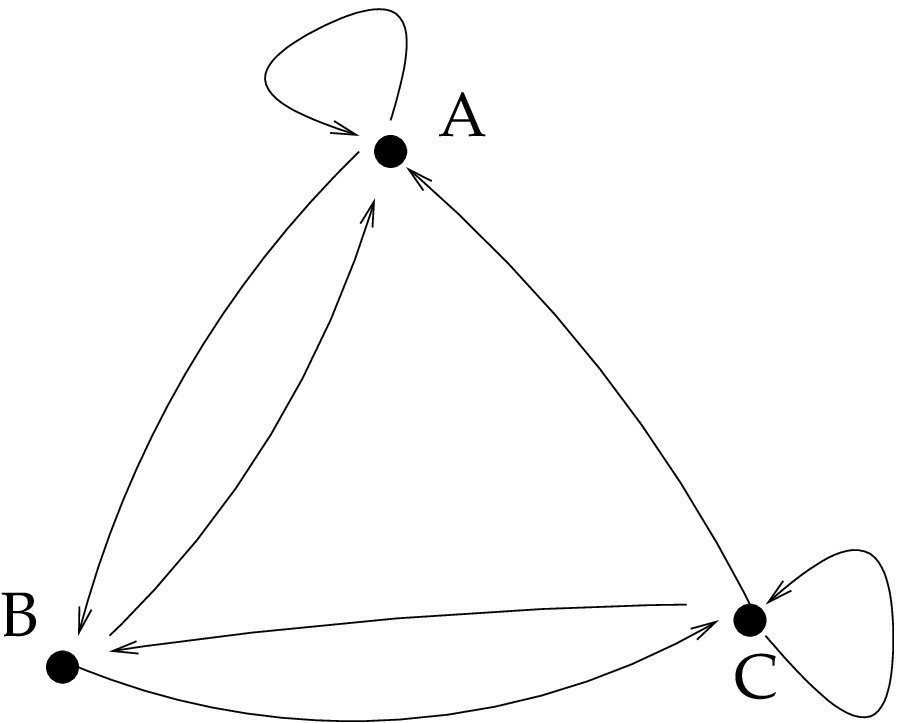}}
\caption{Two examples of transition graphs for which $\rho(\mathbf{Q})\leq 1$ is not a sufficient condition.} \label{fig:Markov_23}
\end{figure}

In the preceding sections we have shown that the classic Kraft inequality is not, in general, a necessary condition for the unique decodability of a constrained sequence, and we have found a necessary condition under this hypothesis. Unfortunately, the found condition is not sufficient and trivial examples show this fact. We note that the only parameter determining the matrix $\mathbf{Q}$ are the length vector $\mathbf{l}$ and the graph associated to the Markov chain, i.e. the state pairs with positive transition probability. Thus, we only consider the transition graphs of the sources without taking into account the value of the transition probabilities. Consider for example a source with three symbols $A,B$ and $C$ with transition graph as shown in fig. \ref{fig:Markov_2}. It is easy to see that if $\mathbf{l}=[1,1,1]$ then $\rho(\mathbf{Q})=1$; anyway, it is clearly impossible to decode the sequences of the source if we assign only one bit to every symbol. In general, we may consider that it is not possible to have a decodable code with more than $2^i$ codewords of length $i$, because otherwise even the initial state cannot be recovered. Anyway, still imposing this additional condition does not suffice. Take for example a code with $\mathbf{l}=[1,1,2]$ for a source with transition graph as shown in fig. \ref{fig:Markov_3}; we have $\rho(\mathbf{Q})<1$, only two codewords of 1 bit and one codeword of 2 bits, but still a decodable code with those lenghts does not exist ($A\to 0$ imposes $B\to 1$, and consequently $C\to 11$, but so $BCB$ and $CC$ have the same code).

The above examples show that the question of finding a sufficient condition for the unique decodability of codes for constrained sequences appears to be more complicated than with unconstrained sequences. A positive fact is that it is possible to extend the Sardinas Patterson (SP) test (\cite{sardinas-patterson-1953}) to the case of our interest. Given a set of codewords, the SP test allows to establish in a finite number of steps if the code is uniquely decodable for unconstrained sequences. Here we modify the classic algorithm for the case of constrained ones. The generalization is straightforward so that we do not give here a formal proof, as it would merely be a rewriting of that for the classic SP test, for which we refer the reader to \cite[th. 2.2.1]{ash-book}.

Suppose our source symbol set is $\mathcal{X}=\{1,2,\ldots,n\}$ and let us call $W=\{W_i\}_{i=1,\ldots,n}$ the set of associated codewords. For $i=1,2,\ldots,n$ we call $F_i=\{W_j|{P}_{ij}>0\}$ the subset of $W$ containing all codewords that can follow $W_i$ in a source sequence. We construct a sequence of sets $S_1,S_2,\ldots$ in the following way. To form $S_1$ we consider all pairs of codewords of $W$; if a codeword $W_i$ is a prefix of another codeword $W_j$, i.e. $W_j=W_iA$ we put the suffix $A$ into $S_1$. In order to consider only the possible sequences, we have to remember the codewords that have generated every suffix; thus, let us say that we mark the obtained suffix $A$ with two labels, thus we indicate it with $_iA_j$. We do this for every $i$ and $j$. Then, for $n>1$, $S_n$ is constructed by comparing elements of $S_{n-1}$ and elements of $W$; for a generic element $_lB_m$ of $S_{n-1}$ we consider the subset $F_l$ of $W$:
\begin{itemize}
\item[a)] If a codeword $W_k\in F_l$ is equal to $_lB_m$ the algorithm stops and the code is not decodable,
\item[b)] if $_lB_m$ is prefix of a codeword $W_r={_lB}_m C$ we put the labelled $_mC_r$ suffix into $S_n$,
\item[c)] if if instead a codeword $W_s$ is prefix of $_lB_m=W_sD$, we place the labelled suffix $_sD_m$ into $S_n$.
\end{itemize}
The code is uniquely decodable if and only if item a) is never reached. Note that the algorithm can be stopped after a finite number of steps; there are in fact only a finite number of possible different sets $S_i$ and so the sequence $S_i,i=1,2,\ldots$ is either finite or periodic. We note that the code is \emph{finite delay} uniquely decodable if the sequence $S_i$ is finite and \emph{infinite delay} uniquely decodable if the sequence is periodic.

\begin{figure}
\centering
\subfigure[\label{fig:Markov_4a}]{\includegraphics[width=3.5cm]{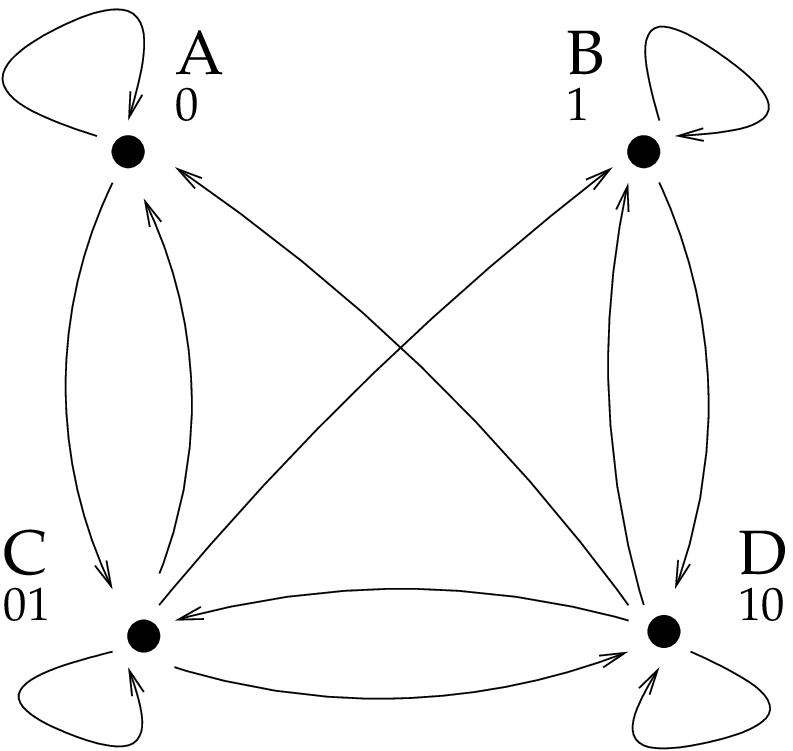}}
\hspace{20pt}
\subfigure[\label{fig:Markov_4b}]{\includegraphics[width=3.5cm]{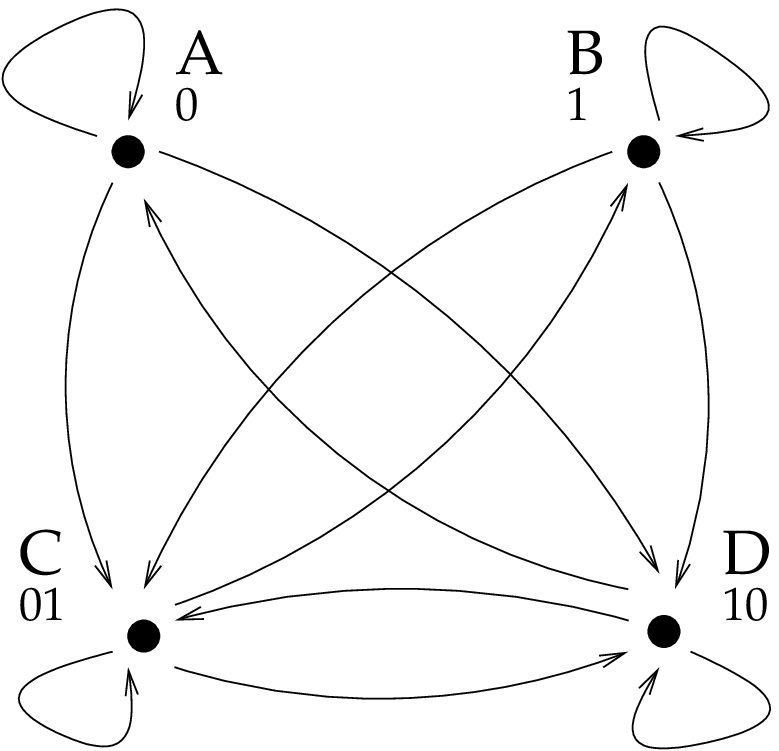}}
\caption{Two examples of transition graphs for which $\rho(\mathbf{Q})\leq 1$ is a sufficient condition.} \label{fig:Markov_4}
\end{figure}

As an example of SP test for constrained sequences we consider the transition graphs shown in fig. \ref{fig:Markov_4}. For both cases we use codewords 0, 1, 01 and 10 for $A,\ B,\ C$ and $D$ respectively. For the graph of fig. \ref{fig:Markov_4a} we obtain $S_1=\{_A1_C, {_B}0_D\}$, $S_2=\emptyset$. Thus the code is finite delay uniquely decodable and we can indeed verify that we need to wait at most two bits for decoding a symbol. For the graph of fig. \ref{fig:Markov_4b}, instead, we have $S_1=\{_A1_C, {_B}0_D\}$, $S_2=\{_C0_D,{_D}1_C\}$, $S_3=S_2$ and then $S_i=S_2$ for every other $i>3$. So, the code is still uniquely decodable but with infinite delay; in fact it is not possible to distinguish the sequences $ADDD\cdots$ and $CCC\cdots$ until they are finished, so that the delay may be as long as we want.

\section{Entropy rate and average lengths}
In the second section we have seen an example of Markov chain that is efficiently encoded with a non prefix-free code. Here we note that the same thing happens with the sources of fig \ref{fig:Markov_4} with the indicated codewords if the transition probability matrix associated are
\begin{equation}\label{eq:Markov_P4}
\mathbf{P}=\left[
\begin{array}{cccc}
1/2 & 0 & 1/2 & 0\\
0 & 1/2 & 0 & 1/2\\
1/4 & 1/4 & 1/4 & 1/4\\
1/4 & 1/4 & 1/4 & 1/4
\end{array}
\right]
\end{equation}
and
\begin{equation}
\mathbf{P}=\left[
\begin{array}{cccc}
1/2 & 0 & 1/4 & 1/4\\
0 & 1/2 & 1/4 & 1/4\\
0 & 1/2 & 1/4 & 1/4\\
1/2 & 0 & 1/4 & 1/4
\end{array}
\right]
\end{equation}
respectively, and the initial state is uniformly distributed. For this sources the entropy of the first $k$ symbols is $(3k+1)/2$ while our code uses on average $3k/2$ bits. Thus we cannot say that, even for stationary ergodic processes\footnote{Note that in this paper we have considered only stationary processes; for nonstationary processes there may be still more surprising results.}, the minimum average length of the code for the first $k$ symbols is greater than or equal to their entropy. The only thing we can say is about the entropy rate. 

Given a Markov source $X_1X_2X_3\cdots$ with transition matrix $\mathbf{P}$, let ${\boldmath{\mu}}=[\mu_1,\mu_2,\ldots,\mu_n]$ be the stationary distribution, and $\mathcal{H}$ the  entropy rate. Using the asymptotic equipartition property for ergodic sources (Shannon-McMillan theorem, \cite{mcmillan-aep}), we deduce that for every uniquely decodable code we must use at lest $\mathcal{H}$ bits per symbol on average and thus $\mathbf{l} \cdot \mu^T \geq \mathcal{H}$. Note that this does not imply that $E[l(X_1)]\geq H(X_1)$, nor, in general, that for some fixed $n$ we have $E[l(X_1,X_2,\ldots,X_k)]\geq H(X_1,X_2,\ldots,X_k)$ (see \cite[Th. 5.4.2]{cover-thomas-book}), and our examples precisely show that in fact this is not true. The point is that for sources with memory we can use codes that do not respect the classic Kraft inequality and thus the inequality $E[l(X)]\geq H(X)$ cannot be proved. Thus, we should be careful in formulating the converse theorem for variable length codes.

Anyway, the examples shown in this and in preceding sections leave many open questions. We have shown examples of transition graphs with associated codeword lengths such that the obtained code is uniquely decodable. For every one of these graphs we have shown that there is a transition probability matrix $\mathbf{P}$ such that the entropy rate of the source exactly equals the average length per symbol of our code. We should ask if this is only a coincidence or if it is the rule. Furthermore, we may ask what happens in terms of efficiency with our code for a generic matrix $\mathbf{P}$ on a given graph; is it possible to find an optimal codeword assignment as we do with Huffman codes in the classic framework? Again, what happens if we consider blocks of more than one symbols? We clearly obtain a different transition graph and the efficiency cannot be lower; but is there a gain in some cases or not?

\section{Some considerations}
It is interesting to consider the proposed coding approach from the point of view of the encoding complexity. Consider for example the stationary Markov chain with transition matrix of eq. (\ref{eq:Markov_P4}) and uniform initial distribution. Note that it is possible to encode the first $k$ symbols of the source using exactly as many bits as the entropy of those symbols using Huffman codes. Two bits are used for the first symbol, which correspond to its entropy. Then, at every step we use a different Huffman code, depending on the preceding step, and use on average $H(X_{k+1}|X_k)=\mathcal{H}=3/2$ bits. Note that this technique actually fits with the conditional entropy idea in the sense that it really encodes each symbol given the preceding one. This implies that the encoder must trace the state of the source and choose the code for the new symbol. On the contrary, the non prefix-free codeword assignment indicated in fig. \ref{fig:Markov_4} allows a very simple encoding phase, as there is a fixed mapping from symbols to code bits, with the same (slightly better) compression performance. The point is that we are making a different use of the decoder knowledge about possible transitions. Note that, even for the Huffman code, we are supposing that the decoder exactly knows what transitions are possible and what are not. The difference is that with the non prefix-free code we are making the decoder more active. This relates the presented idea to other developed coding paradigms. We should note in fact, that in practice the proposed approach was already used in other contexts. One of the oldest examples may be that of modulo-PCM codes (\cite{modulo-PCM}) for numerical sequences; here only the modulo-4 value of every sample is encoded, leaving to the decoder the task of understanding the original value using its knowledge on the memory of the source.
In that case the used code is even a singular code\footnote{In our examples here we have considered only non singular codes with singular extension. The non singularity of the code is required in our setting where we want to be able to decode a sequence composed of one single symbol.} but under certain hypothesis this does not affect the decodability. 
Similar ideas are then used in the recently reemerged theme of distributed source coding (see for example \cite{pradhan-magazine} and references therein). Let us consider for a moment the problem of noiseless separated coding of dependent sequences. The well known Slepian-Wolf theorem (\cite{slepian-wolf-1973}) says that two correlated memoryless sources, $X$ and $Y$, can be separately lossless coded at rates $R(X)$ and $R(Y)$ respectively when jointly decoded, if $R(X)\geq H(X|Y)$, $R(Y)\geq H(Y|X)$ and $R(X)+R(Y)\geq H(H,Y)$. Cover extended the result to the case of general ergodic sources in \cite{cover-slepianwolf}. Roughly speaking, the used encoding process at every encoder consists on considering large blocks of symbols; the set of all such blocks is split into disjoint bins and only the index of the bin that contains the extracted block is encoded. At the decoder the original block for both sequences $X$ and $Y$ is recovered by extracting from the pair of specified bins the only pair of jointly typical blocks. It is interesting to note that this encoding technique actually uses singular codes in order to achieve compression, leaving to the decoder the task of disambiguating, based on the joint statistic of the two sources. 

The same idea is now being used, in order to shift the complexity from the encoder to the decoder, in what may be called ``source coding based on distributed source coding principles''. Special attention in this field is being payed to the case of video coding (see for example \cite{puri-PRISM,Girod-DVC}). In this contest the memory of the video source is not exploited in the encoding phase but in the decoding one. Again, roughly speaking, in the encoding phase no motion compensation and prediction is applied and singular codes are used for the data compression task. In the decoding phase, on the contrary, the memory of the source is exploited in order to remove ambiguity by using motion compensation. It is interesting to see that practical architectural specification for this video coding techniques have been presented only in recent years, even if the idea of such an approach to video coding was already patented by Witsenhausen and Wyner in late '70s (\cite{WW-patent}). Moreover, it is also interesting to note that there is not much difference in this approach with respect to the idea behind the modulo-PCM coding above mentioned.

As a final comment on the general problem of finding variable length non prefix-free codes for a given constrained sequence, we note that there are many connections with the area of coding for constrained channels (see for example \cite{symbolic-dynamics-book}). The relation between the transition graphs of our sources and the possible codeword assignments may find interesting counterparts (and maybe answers) in that field, where graph theory has already been successfully applied.
\bibliographystyle{IEEEbib}
\bibliography{bibeit}

\end{document}